\documentclass[%
reprint,
 aps,
 pra,
]{revtex4-1}

\usepackage{amssymb,amsmath,amsfonts,bm}
\usepackage{graphics,graphicx}
\usepackage{dcolumn}
\usepackage{float}
\usepackage{bm}
\usepackage{parskip}
\usepackage{amsmath}
\newtheorem{theorem}{Theorem}

\usepackage{multirow}
\usepackage{fancyhdr}
\usepackage{wasysym}
\usepackage{url}
\usepackage{color}
\usepackage[colorlinks,linkcolor=blue,anchorcolor=blue,urlcolor=blue,citecolor=blue]{hyperref}

\usepackage{ulem}

\begin{document}
\title{Linear Response Theory of Entanglement Entropy}

\author{Yuan-Sheng Wang}
\affiliation{Department of Physics, Southern University of Science and Technology, Shenzhen 518055, China}
\affiliation{School of Physical Sciences, University of Science and Technology of China, Hefei 230026, China}

\author{Teng Ma}%
 \affiliation{Beijing Academy of Quantum Information Sciences, Beijing 100193, China}

\author{Man-Hong Yung}\email{yung@sustech.edu.cn}
\affiliation{Institute for Quantum Science and Engineering, and Department of Physics, Southern University of Science and Technology, Shenzhen 518055, China}

\date{\today}
\begin{abstract}
  Linear response theory (LRT) is a key tool in investigating the quantum matter, for quantum systems perturbed by a weak probe, it connects the dynamics of experimental observable with the correlation function of unprobed equilibrium states.
  Entanglement entropy(EE) is a measure of quantum entanglement, it is a very important quantity of quantum physics and quantum information science.  
  While EE is not an observable, 
  developing the LRT of it is an interesting thing. 
  In this work, we develop the LRT of von Neumann entropy for an open quantum system.
  Moreover, we found that the linear response of von Neumann entanglement entropy is determined by the linear response of an observable. Using this observable, we define the Kubo formula and susceptibility of EE, which have the same properties of its conventional counterpart.   
  Through using the LRT of EE, we further found that the linear response of EE will be zero for maximally entangled or separable states, this is a unique feature of entanglement dynamics.   
  A numerical verification of our analytical derivation is also given using XX spin chain model.
  The LRT of EE provides a useful tool in investigating and understanding EE. 
\end{abstract}

\maketitle

\section{\label{sec:intro}Introduction}

Entanglement provides fundamental insight into quantum physics\cite{PhysicsPhysiqueFizika.1.195,PhysRevLett.70.1895} and quantum information science\cite{Bennett2000}. For example, in condensed matter physics, entanglement provides a unique tool for characterizing quantum mechanical many-body phases and new kinds of quantum order\cite{PhysRevLett.96.110405,PhysRevLett.96.110404}.
As a bridge linking quantum statistical mechanics with quantum information, entanglement has been exploited to establish efficient methods for examining many-body interacting systems\cite{Osborne2002}. 
Deep understanding of the entanglement decoherence is also expected to insights into the quantum measurement and the quantum-classical transition\cite{GISIN1996151,PhysRevA.69.052105,PhysRevA.69.052106}.
Besides the fundamental physical meaning of entanglement, it is also a resource required for many quantum technologies such as quantum communication, quantum sensing\cite{RevModPhys.89.035002}, quantum metrology\cite{PhysRevLett.96.010401,Giovannetti2011,PhysRevLett.109.233601}, etc. 
The ability to generate, manipulate and detect entanglement\cite{Zeilinger_1998,Tittel_1998,Deutsch_1998,DiVincenzo_1998} are bases of the whole field of quantum computing.

The interest in the properties of entanglement has also been extended to understanding its dynamical behaviour, 
which is a powerful perspective in understanding non-equilibrium quantum dynamics\cite{PhysRevLett.99.160502,PhysRevB.98.035118,zhu2019entanglement}.
For instance, the dynamics of entanglement has recently been realized as a useful probe in studying ergodicity
and its breakdown in quantum many-body systems\cite{PhysRevB.77.064426,PhysRevLett.109.017202,PhysRevLett.111.127205,PhysRevB.95.094302}. 
Although great progress has been made in this field, to our knowledge, however, 
there is still a lack of a general tool in investigating how entanglement entropy (EE) response to an external perturbation.


Conventional linear response theory (LRT) is a general and powerful tool in studying quantum matters\cite{PhysRevLett.70.65,PhysRevB.44.1646,PhysRevA.30.1185,PhysRevLett.58.1861,HANGGI1982207,PhysRevB.54.16487}, 
because it connects the dynamical response of a quantum system to an external probe with the correlation functions of the unperturbed equilibrium state. 
While the conventional LRT is applied to observable, which is associated with a Hermitian operator,
as EE is not an observable, 
finding the LRT for it seems to be not only interesting but also useful.
 

In this paper, we develop an LRT of von Neumann entropy for an open quantum system which is subject to external perturbation.
We found that the LRT of von Neumann entropy to external perturbation is given by the linear response of an observable, using this observable we define the Kubo formula and susceptibility of EE which have the same properties of its conventional counterpart. 
Using the LRT of EE, we demonstrate that there is no linear response of von Neumann entropy for non-degenerate systems which are initially at separable states and maximally entangled states. 
To verify our analytical result, we numerically solve the dynamics of von Neumann entropy of a subsystem in a XX spin chain model.
The LRT of EE provides a useful tool in understanding and investigating the dynamics of entanglement.

The paper is organized as follows. In Sec.\ref{sec:review}, we briefly review the conventional LRT.  In Sec.\ref{sec:linear} we derive the LRT of von Neumann entropy, define the Kubo formula and the susceptibility of EE, discuss properties and applications of the Kubo formula and the susceptibility of EE. In Sec.\ref{sec:numerical} we give a numerical verification of our analytical results using XX spin chain model. The concluding Sec. \ref{sec:conclusion} summarizes the findings. Technical details of some calculations are presented in the appendix.

\section{\label{sec:review}Review of conventional linear response theory}
\subsection{\label{subsec:qlrf}The Kubo formula}
Consider a quantum system which is weakly coupled to an external perturbation $F(t)\hat{H}_{1}$ at $t=t_{0}$ with time dependence $F(t)$. The Hamiltonian for such system is given by  
\begin{align}
    \hat{H}(t)=\hat{H}_{0}+F(t)\hat{H}_{1}
    \label{hamiltonian}
\end{align}
where $\hat{H}_{0}$ is the Hamiltonian of unperturbed system. 
In interaction picture (we denote an operator or density matrix in interaction picture through adding "I" to its subscript),
the average of any operator $\hat{O}_{I}(t)$ over state $\rho_{I}(t)$ reads
\begin{align}
\langle\hat{O}_{I}(t)\rangle&\equiv\text{tr}[\rho_{I}(t)\hat{O}_{I}(t)]
\end{align}
where $\rho_{I}(t)=\hat{U}_{I}(t,t_{0})\rho_{I}(t_{0})\hat{U}_{I}^{\dagger}(t,t_{0})$
with the time evolution operator $\hat{U}_{I}(t,t_{0})=\mathcal{T}\exp[-i\int_{t_{0}}^{t}dt^{\prime}F(t^{\prime})\hat{H}_{1,I}(t^{\prime})]$.
If $F(t)$ is so small such that $F(t)\hat{H}_{1}$ is much smaller than $\hat{H}_{0}$, 
we can describe $\langle\hat{O}_{I}(t)\rangle\equiv\text{tr}[\rho_{I}(t)\hat{O}_{I}(t)]=\text{tr}[\rho(t)\hat{O}(t)]$ by performing an expansion in powers of $F(t)$, 
working to the first order solution to $\hat{U}_{I}(t)$ (we assume $\hbar=1$ hereafter):
\begin{align}
    \delta\langle\hat{O}_{I}(t)\rangle= \int_{t_{0}}^{\infty}dt^{\prime}R(t,t^{\prime})F(t^{\prime}))+\mathcal{O}(F^{2})
    \label{kubo_f1}
\end{align}
where $\delta\langle\hat{O}_{I}(t)\rangle\equiv\langle\hat{O}_{I}(t)\rangle_{F}-\langle\hat{O}_{I}(t)\rangle_{F=0}$ and $R(t,t^{\prime})=-i\theta(t-t^{\prime}))\langle[\hat{O}_{I}(t),\hat{H}_{1,I}(t^{\prime})]\rangle_{0}$, 
with $\langle\cdot\rangle_{0}\equiv\text{tr}\left[\cdot\rho(t_{0})\right]$ an average with respect to the initial state $\rho(t_{0})$. 
The real function $R(t,t^{\prime})$ is the well-known Kubo formula of observable $\hat{O}$ for system which is subject to a perturbation $F(t)\hat{H}_{1}$, which is also referred to as the linear response function. 
It is readily to see that the Kubo formula can be written in terms of two-point correlation function $C(t,t^{\prime})\equiv\langle\hat{O}_{I}(t)\hat{H}_{1,I}(t^{\prime})\rangle_{0}$, 
if $\rho(t_{0})$ is an equilibrium state, the correlation function  provides a measure of equilibrium fluctuations in the system, 
therefore, it is revealed by Eq.\eqref{kubo_f1} that although the response of an equilibrium system to external perturbation should move the system away from equilibrium, 
if the perturbation is weak enough, the response is dictated by the equilibrium fluctuations.  
We may conclude that the knowledge of the equilibrium state is useful in predicting the behaviour of nonequilibrium process.

\subsection{Some properties and applications of the Kubo formula}
\textit{Stationary}.
If the initial state is equilibrium state, its density matrix then commutes with $\hat{H}_{0}$, 
inserting $\hat{O}_{I}(t)=e^{i\hat{H}_{0}t}\hat{O}e^{-i\hat{H}_{0}t}$ 
and $\hat{H}_{1,I}(t)=e^{i\hat{H}_{0}t^{\prime}}\hat{H}_{1}e^{-i\hat{H}_{0}t^{\prime}}$ into the expression of correlation function and using the cyclic property of trace, we have 
\begin{align}
    C(t,t^{\prime})&=\langle\hat{O}_{I}(t-t^{\prime})\hat{H}_{1}\rangle
\end{align}
that is, the equilibrium state correlation function is invariant under time translations: $C(t,t^{\prime})=C(t-t^{\prime})$,  
which results in translation invariant Kubo formula: $R(t,t^{\prime})=R(t-t^{\prime})$.

\textit{Causal}. The system should not respond before the external perturbation is applied, therefore $R(t)=0$ for $t<0$, this is enforced by the step function in the expression of $R(t)$. 
The causality will result in analyticity of $\chi(\omega)$ in the upper half complex plane, we will show it in the following. 
The Fourier expansion of $R(t)$ reads
\begin{align}
    R(t)=\frac{1}{2\pi}\int_{-\infty}^{\infty}\chi(\omega)e^{-i\omega t}d\omega
\end{align}
if $t<0$ we can perform the integral by completing the contour in the upper half plane, the complementary integral has to be zero since the real part of the exponent $-i(i|\omega_{i}|)\cdot(-|t|)\rightarrow -\infty$ along the contour in the upper half plane, with $\omega_{i}$ the imaginary part of $\omega$. On the other hand, according to Cauchy's residue theorem, the integral along the contour is given by the sum of the residues inside the contour. So there is no poles of $\chi(\omega)$ for $\text{Im}\chi(\omega)>0$, which means that $\chi(\omega)$ is analytic in the upper half plane.   
\ \\
\textit{Kramers-Kr\"onig relation}. Because $\chi(\omega)$ is analytic in the upper half complex plane, it is found that the imaginary and real parts of $\chi(\omega)$ is not independent, 
they are connected by the Kramers-Kr\"onig relation\cite{PhysRev.104.1760,deL.Kronig:26,Kramers1927}:
\begin{align}
    \text{Re}[\chi(\omega)]&=\frac{1}{\pi}\mathcal{P}\int_{-\infty}^{\infty}\frac{\text{Im}[\chi(\omega)]}{\omega^{\prime}-\omega}d\omega^{\prime}\\
    \text{Im}[\chi(\omega)]&=-\frac{1}{\pi}\mathcal{P}\int_{-\infty}^{\infty}\frac{\text{Re}[\chi(\omega)]}{\omega^{\prime}-\omega}d\omega^{\prime}.
\end{align}
where $\mathcal{P}$ denotes the Cauchy principal value. Using Kramers-Kr\"onig relation, the response function can be reconstructed given just real or imaginary part.

\textit{Susceptibility}. 
According to the stationary property, for equilibrium initial state, the Kubo formula is time-translation invariant: $R(t,t^{\prime})=R(t-t^{\prime})$, 
Eq.\eqref{kubo_f1} becomes a convolution, its Fourier transform reads
\begin{align}
    \langle\delta\hat{O}_{I}(\omega)\rangle=\chi(\omega)F(\omega)+\mathcal{O}(F^2)
    \label{f_input_output}
\end{align}
where $F(\omega)$ is the Fourier transforms of $F(t)$, $\chi(\omega)$ is the Fourier transform of Kubo formula and is oftern referred to as generalized susceptibility.
We learn from Eq.\eqref{f_input_output} that the response of observable $\hat{O}$ to time-dependent perturbation $F(t)\hat{H}_{1}$ is 'local' in frequency space, 
this is a feature of equilibrium linear response. 
Thanks to this locality, we can directly infer the generalized susceptibility $\chi(\omega)$ of $\hat{O}$ for a known perturbation $F(t)\hat{H}_{1}$ through measuring the time evolution of $\langle\hat{O}_{I}(t)\rangle$. 
The susceptibility of some observable to certain perturbation is essential in investigating the properties of equilibrium systems (e.g. classifying different materials using magnetic susceptibility).    
Therefore the conventional LRT is an important tool in investigating the properties of unknown systems.


\textit{Fluctuation-dissipation relation}. Denoting the imaginary part of generalized susceptibility $\chi(\omega)$ as $\chi^{\prime\prime}(\omega)$, it can be written as 
\begin{align}
    \chi^{\prime\prime}(\omega)=\frac{1}{2i}\int_{-\infty}^{\infty}dt e^{i\omega t}[R(t)-R(-t)]
\end{align}
which is not invariant under time-reversal $t\rightarrow -t$. 
Hence, $\chi^{\prime\prime}(\omega)$ is called the dissipative or absorptive part of the generalized susceptibility.
The fluctuation–dissipation theorem says that the dissipation effects are determined by natural fluctuation in thermal equilibrium state. To be more specific, for canonical ensemble with the inverse temperature $\beta$, $\rho =e^{-\beta \hat{H}_{0}}/Z$, the imaginary part of susceptibility $\chi^{\prime\prime}(\omega)$ which describe the dissipation and (the Fourier transform of) the correlation function $\tilde{C}(\omega)$ are connected by the Fluctuation-dissipation relation: 
\begin{align}
   \tilde{C}(\omega)=-2[n_{B}(\omega)+1]\chi^{\prime\prime}(\omega)
\end{align}
where $n_{B}(\omega)=1/(e^{\beta\omega}-1)$ is the Bose-Einstein distribution function.

\section{\label{sec:linear}The Linear Response Theory of Entanglement Entropy}
\subsection{The Kubo formula of EE}
Be similar to the conventional case, 
supposing at time $t=t_{0}$, a time-dependent perturbation $F(t)\hat{H}_{1}$ is applied to a composite system which consists of subsystems A and B, its Hamiltonian has the same form as \eqref{hamiltonian}.
Solving the dynamics of this composite system using perturbative method (see appendix \ref{derivative_1} for detail), we have 
\begin{align}
  \delta\rho(t)=-i\int_{t_{0}}^{t}d\tau[\hat{H}_{1,I}(\tau-t),\rho_{0}]F(\tau)+\mathcal{O}(F^{2}), 
  \label{dyna_state}
\end{align}
where $\rho_{0}\equiv \rho(t_{0})$, $\delta\rho(t)\equiv\rho(t)-\rho_{0}$, 
and $\hat{H}_{1,I}(t)=\hat{U}_{0}^{\dagger}(t,t_{0})\hat{H}_{\text{1}}\hat{U}_{0}(t,t_{0})$,
with $\hat{U}_{0}(t,t_{0})=\exp[-i\hat{H}_{0}(t-t_{0})]$.

The entanglement entropy of subsystem A is given by $S_{A}(t)=-\text{tr}[\rho_{A}(t)\ln\rho_{A}(t)]$, 
with $\rho_{A}(t)=\text{tr}_{B}[\rho(t)]$.
For infinitesimal change $\delta\rho_{A}(t)\equiv\rho_{A}(t)-\rho_{A}(t_{0})$, 
the time-dependent change in entanglement entropy reads\cite{PhysRevA.99.062303}: 
$\delta S_{A}(t)\equiv S_{A}(t)-S_{A}(t_{0})\approx-\text{tr}_{A}\left[\delta\rho_{A}(t)\ln\rho_{A}(t)\right]$, 
furthermore, we can expand the operator $\ln\rho_{A}(t)$ in terms of $\delta\rho_{A}(t)$ as 
$\ln\rho_{A}(t)=\ln\rho_{A}(t_{0})|+\mathcal{O}(\delta\rho_{A})$, 
then to the first order in $\delta\rho_{A}(t)$, we have
\begin{align}
  \delta S_{A}(t)=-\text{tr}_{A}\left[\delta\rho_{A}(t)\ln\rho_{A}(t_{0})\right]+\mathcal{O}(\delta\rho_{A}).
  \label{dyna_s}
\end{align}
What's interesting about this equation is that
the right-hand side of \eqref{dyna_s} has the form of the change in the expectation value of an Hermitian operator $\ln \rho_{A}(t_{0})$.
Similar result can be seen in Ref. \cite{PhysRevX.7.021003}.


Inserting Eq. \eqref{dyna_state} into Eq. \eqref{dyna_s}, 
we obtain the time evolution of EE under weak driving (see appendix \ref{derivative_2} for detail):
\begin{align}
    \delta S_{A}(t)=\int_{t_{0}}^{\infty}R_{E}(t,\tau)F(\tau)d\tau+\mathcal{O}(F^2)
    \label{convolution}
\end{align}
where $R_{E}(t,\tau)$ is given by the following expression
\begin{align}
    R_{E}(t,\tau)=-i\theta(t-\tau)\langle[\hat{s}_{A},\hat{H}_{1,I}(\tau-t)]\rangle_{0}. 
    \label{kubo-f}
\end{align}
where $\hat{s}_{A}\equiv-\ln \rho_{A}(t_{0})$, 
Comparing Eq. \eqref{convolution} and  Eq. \eqref{kubo-f} with Eq. \eqref{kubo_f1}, 
we found that to first order in $F(t)$, the change in EE is equivalent to the change in the expectation of observable $\langle\hat{s}_{A}(t)\rangle$:
\begin{align}
\delta S_{A}(t)\approx\delta\langle \hat{s}_{A}(t)\rangle 
\label{equivalent}
\end{align}
Eq. \eqref{convolution}, Eq. \eqref{kubo-f} and Eq. \eqref{equivalent} are key results of this work,
they revealed that: \textit{for equilibrium open system A, the response of EE is determined by the linear response of observable} $\hat{s}_{A}=-\ln \rho_{A}(t_{0})$. 
Eq. \eqref{kubo-f} has the same form as conventional linear response function, 
it describes how EE and the observable $\hat{s}_{A}$ response to weak external force $F(t)\hat{H}_{1}$, we refer to $R_{E}(t,\tau)$ as the Kubo formula of EE.    

\subsection{Properties and applications of the EE Kubo formula}

The Kubo formula of EE is also the Kubo formula of an observable, 
this fact implies that $R_{E}(t,\tau)$ has the same properties of conventional linear response function. 

Like conventional LRT, 
if $[\rho_{0},\hat{H}_{0}]=0$, the EE Kubo formula will be \textit{stationary}: $R_{E}(t,\tau)=R_{E}(t-\tau)$,
then the Fourier transform(FT) of Eq.\eqref{convolution} reads
\begin{align}
  \delta S_{A}(\omega)=\chi_{E}(\omega)F(\omega)+\mathcal{O}(F^2) \label{LR_f},
\end{align}
where $\chi_{E}(\omega)$ and $F(\omega)$ are FT of $R_{E}(t)$ and $F(t)$ respectively. 
Eq. \eqref{LR_f} has the same form as \eqref{f_input_output} except for that $S_{A}(t)$ is not an expectation value of an observable.
$R_{E}(t)$ and $\chi_{E}(\omega)$ also satisfy properties such as \textit{causality}, \textit{analyticity} of $\chi_{E}(\omega)$ in the upper-half plane and there is a \textit{Kramers-Kr\"oning relation} between real and imaginary part of $\chi_{E}(\omega)$.

\textit{Entanglement susceptibility}. We refer to $\chi_{E}(\omega)$ as entanglement susceptibility.
According to Eq. \eqref{convolution}, the time evolution of EE can be determined by measuring an observable $\hat{s}_{A}(t)$. 
$F(t)$ is normally known to the one who applies the perturbation, 
we thus can infer the Kubo formula $R_{E}(t)$ from \eqref{LR_f} and hence the entanglement susceptibility $\chi_{E}(\omega)$. 
The Kubo formula is independent of $F(t)$, for a given $\hat{H}_{1}(t)$, once we determine its Kubo formula, 
we may predict the response for any $F(t)$. Although $\hat{s}_{A}$ is hard to measure, for a system with few degrees of freedom, 
(e. g., a spin-1/2) we can evaluate it using state tomography.

\textit{Canonical ensemble EE linear response}  
If the initial state of subsystem A is described by canonical ensemble $\rho_{A}(t_{0})=e^{-\beta\hat{H}_{A}}/Z_{A}$ with inverse temperature $\beta=1/k_{B}T$, 
$\hat{H}_{A}$ is the Hamiltonian of subsystem A 
and $Z_{A}$ is the corresponding partition function, 
then $\ln \rho_{A}(t_{0})=-\beta\hat{H}_{A}-\ln Z_{A}$, inserting this equation into \eqref{convolution}, we have
\begin{align}
    -k_{B}T\delta S_{A}(t)=\int_{t_{0}}^{\infty}(1/\beta)R_{E}(t,\tau)F(\tau)d\tau+\mathcal{O}(F^2)
    \label{canonical}
\end{align}
it is readily to see that $(1/\beta)R_{E}(t,\tau)=-i\theta(t-\tau)\langle[\hat{H}_{A},\hat{H}_{1,I}(\tau-t)]\rangle_{0}$, 
which is the Kubo formula of observable $\hat{H}_{A}$ and perturbation $\hat{H}_{1}(t)$, 
$\delta Q_{A}(t)\equiv-k_{B}T\delta S_{A}$ has a physical meaning of the heat change of subsystem A during a process with fixed temperature $T$. 
We can conclude from Eq. \eqref{canonical} that $\delta Q_{A}(t)=\langle\hat{H}_{A}(t)\rangle$ to the first order in $F(t)$, this implies that for an open quantum system which is described by the canonical ensemble, if we know how its internal energy responds to a known perturbation, we then know how its EE responds to the same perturbation. 


\subsection{Exceptional states}
Given an LRT of EE, a natural question arises: the sign of response is determined by the sign of driving, or the sign of $\alpha$,
consider a separable initial state, then $S_{A}(t_{0})=0$, we can always choose an appropriate sign of $\alpha$ such that the linear correction of EE is negative,
that is $\delta S_{A}(t)<0$, then $S_{A}(t)=S_{A}(t_{0})+\delta S_{A}(t)<0$, this is, of course, not true. We may have a similar question for maximally entangled states.
Viewed from another perspective, the EE of product states and maximally entangled states must be extreme values, therefore $\delta S_{A}=0$ for these states. To address these problems, we propose Theorem \ref{t1}.  

\begin{theorem}
 For any product state or maximally entangled state of subsystems A and B which is an eigenstate of composite system A$\cup$B, 
 if a time-dependent perturbation $F(t)\hat{H}_{1}$ (with $F(t)\ll 1$) is applied to the composite system, 
 then the linear response of EE of system A or B must be zero. Moreover, $\delta S_{A/B}(t)=\mathcal{O}(\alpha^{2})$.
 
 \label{t1}
\end{theorem}


\textit{Proof}. We here give the proof of the statement about separable states in Theorem \ref{t1}, the proof about maximally states is similar and we show it in appendix \ref{proof_2}. The basic idea of this proof is to prove that $\ln\rho_{A}(t_{0})$ is diagonal while any diagonal element of $\delta\rho_{A}(t)$ is zero, 
then $\text{tr}_{A}[\delta\rho_{A}(t)\ln \rho_{A}(t_{0})]=0$.

For a composite system consists of two subsystem A and B, if its eigenstates are separable or product states of A and B, 
then the Hamiltonian of this composite system must be an non-interacting Hamiltonian of A and B, 
\begin{align}
  \hat{H}_{0}=\hat{H}_{A}+\hat{H}_{B}
\end{align}
before proceeding, we assume that $\hat{H}_{0}$ is non-degenerate, then both subsystem A and B are non-degenerate, their eigenstates and eigenvalues are given by
\begin{align}
  \hat{H}_{A}|f_{i_{A}}\rangle&=f_{i_{A}}|f_{i_{A}}\rangle \\
  \hat{H}_{B}|g_{j_{B}}\rangle&=g_{j_{B}}|g_{j_{B}}\rangle 
\end{align}
the eigenvectors of $\hat{H}_{0}$ are product states of the eigenvectors of A and B:
\begin{align}  
  \hat{H}_{0}\left(|f_{i_{A}}\rangle\otimes|g_{j_{B}}\rangle\right)=(f_{i_{A}}+g_{j_{B}})|f_{i_{A}}\rangle\otimes|g_{j_{B}}\rangle
\end{align}
denoting $|S_{i_{A},j_{B}}\rangle=|f_{i_{A}}\rangle\otimes|g_{j_{B}}\rangle$, it is readily to see that if the initial state is diagonal with respect to basis $\{|S_{i_{A},j_{B}}\rangle\}$, 
that is $\rho_{0}=\sum_{i_{A},j_{B}}P_{i_{A},j_{B}}|S_{i_{A},j_{B}}\rangle \langle S_{i_{A},j_{B}}|$, 
then the reduced density matrix of subsystem A is given by 
\begin{align}
  \rho_{A}(t_{0})=\text{tr}_{B}[\rho_{0}]=\sum_{i_{A}}\Big(\sum_{j_{B}}P_{i_{A},j_{B}}\Big)|f_{i_{A}}\rangle \langle f_{i_{A}}|
\end{align}
which is diagonal with respect to eigenbasis of $\hat{H}_{A}$, so is $\ln \rho_{A}(t_{0})$.

Given the intial state $\rho _{0}=\sum_{i_{A},j_{B}}P_{i_{A},j_{B}}|S_{i_{A},j_{B}}\rangle \langle S_{i_{A},j_{B}}|$ which is diagonal,
if $[\hat{H}_{0},\rho_{0}]=0$ then
an arbitrary entry of the commutator in Eq.\eqref{dyna_state} is given by $\langle S_{i_{A},j_{B}}|[\hat{H}_{1,I}(t-\tau),\rho_{0}]|S_{k_{A},l_{B}}\rangle=(p_{i_{A},j_{B}}-p_{k_{A},l_{B}})\langle S_{i_{A},j_{B}}|\hat{H}_{1,I}(t-\tau)|S_{k_{A},l_{B}}\rangle$, 
and
\begin{align}
  \langle\delta\rho(t)\rangle_{m,m}&=-i\int_{t_{0}}^{t}d\tau\langle S_{m}|[\hat{H}_{1,I}(t-\tau),\rho_{0}]|S_{m}\rangle F(\tau)\nonumber \\
  &\ \ \ \   +\mathcal{O}(F^2)\nonumber \\
  &=\mathcal{O}(F^2)
\end{align}
where $\langle\delta\rho(t)\rangle_{m,m}\equiv\langle\rho(t)-\rho_{0}\rangle_{m,m}$. The above equation implies that after applying perturbation, any diagonal element of linear correction to initial state is zero,
does this means that any diagonal element of $\delta\rho_{A}(t)$ is zero? 
Consider an arbitary off-diagonal element $|S_{i_{A},j_{B}}\rangle\langle S_{k_{A},l_{B}}|$ of the initial state, 
\begin{align}
  \text{tr}_{B}(|S_{i_{A},j_{B}}\rangle\langle S_{k_{A},l_{B}}|)=\delta_{j_{B},l_{B}}|f_{i_{A}}\rangle\langle f_{k_{A}}|
\end{align}
which does not contribute to the diagonal element of subsystem A with respect to basis ${|f_{i_{A}}}\rangle$. 
As any diagonal element of $\rho_{0}$ is zero, we may conclude that the $\delta\rho_{A}(t)$ is hollow, that is the diagonal entries are all zero.

In summary, if the initial eigenstate is a separable state of A and B, moreover, if $\hat{H}_{A}+\hat{H}_{B}$ is non-degenerate, then $\ln \rho_{A}(t_{0})$ is diagonal,
while $\delta\rho_{A}(t)$ is hollow. Thus the linear response of EE $\delta S_{A}(t)=\mathcal{O}(F^2)$.


\begin{figure}
  \centering
  \includegraphics[width=0.4\textwidth]{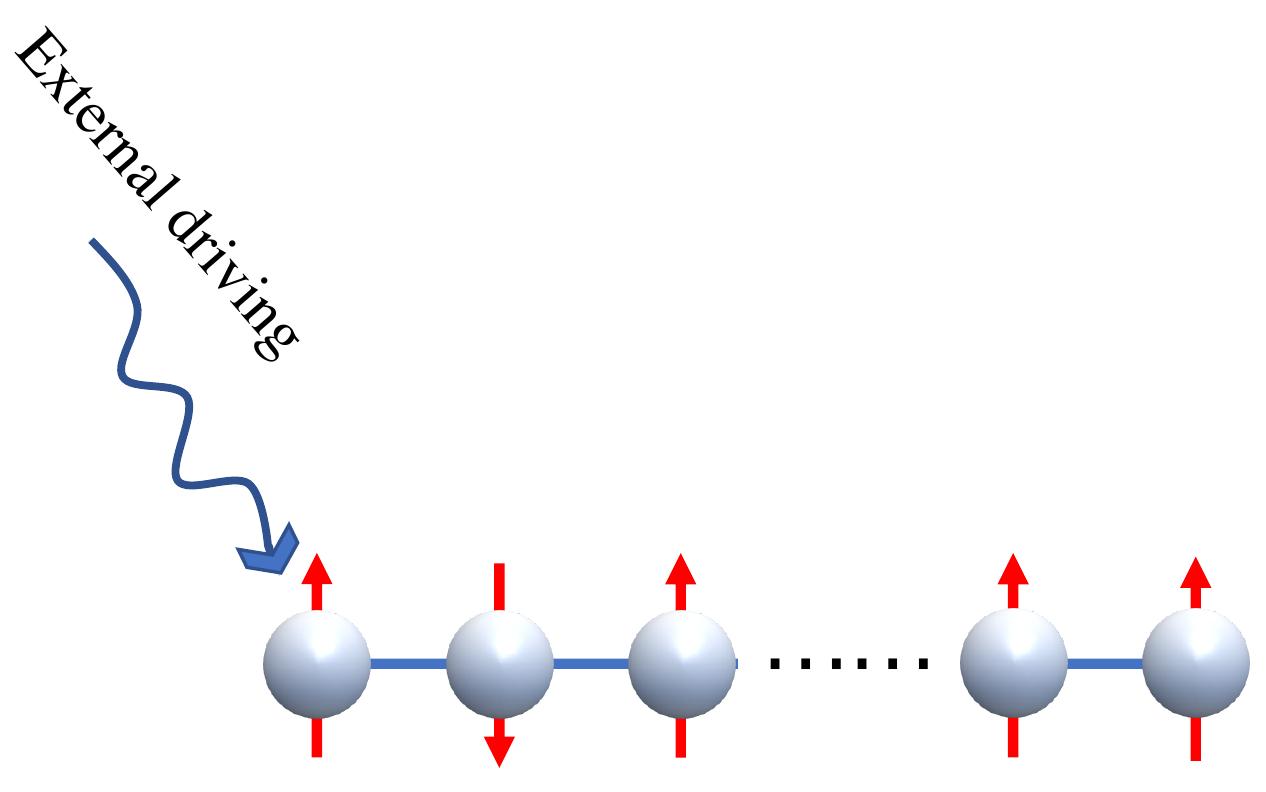}
  \caption{XX spin chain with external driving on the first spin.}
\end{figure}
\begin{figure}
  \centering
  \includegraphics[width=0.4\textwidth]{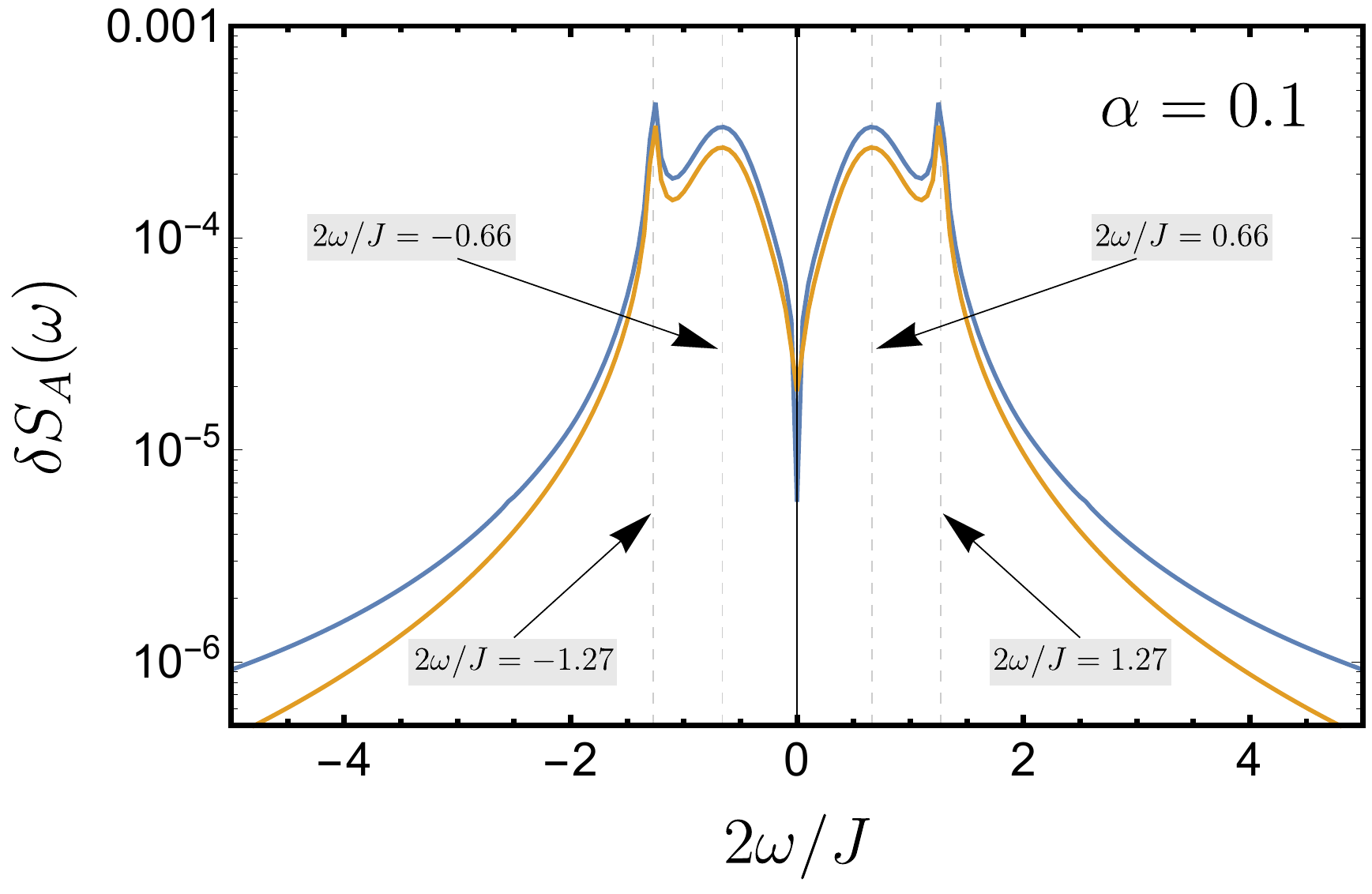}
  \caption{Spectrum of linear response (yellow) and numerical results(blue) for driving amplitude $\alpha=0.1$, other parameters: $L=100$, $J=2.0$, $T=0.5\pi$.}
  \label{comp1}
\end{figure}
\begin{figure}
  \centering
  \includegraphics[width=0.4\textwidth]{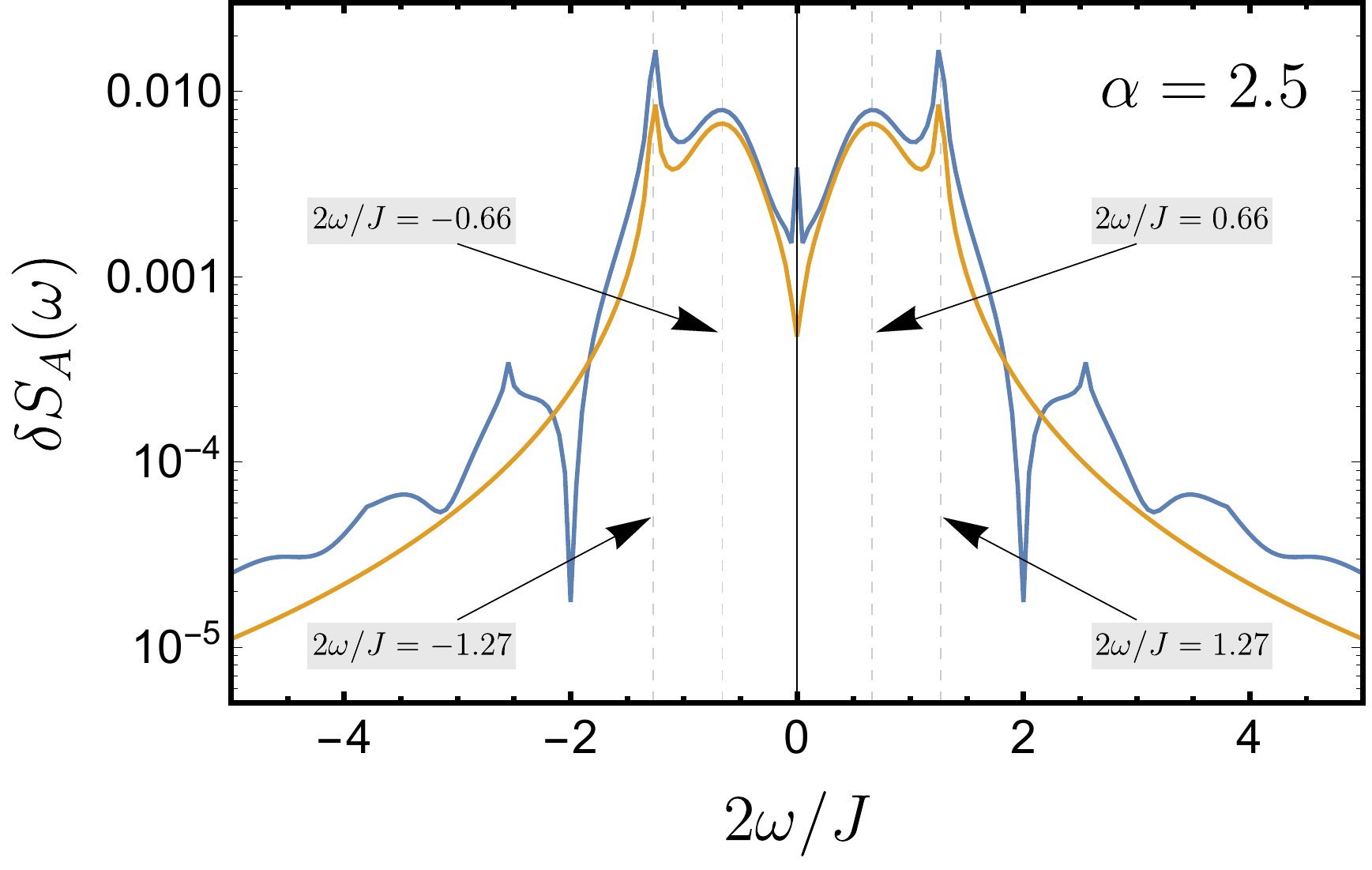}
  \caption{Spectrum of linear response (yellow) and numerical results(blue) for driving amplitude $\alpha=2.5$. Other parameters are the same as Fig. \ref{comp1}.}
  \label{comp2}
\end{figure}

\section{\label{sec:numerical}Linear Response of EE in XX Spin Chain}

To verify our analytical result, we numerically investigate the dynamics of a one-dimension spin chain,
which is composed of $L$ spin-$1/2$ particles coupled via $XX$-type interaction. Hamiltonian of this model 
in the absence of driving reads (we assume $\hbar=1$ hereafter)
\begin{align}
    \hat{H}_{0}=\frac{\lambda}{2}\sum_{j=1}^{L}\hat{\sigma}_{j}^{z}+\frac{J}{2}\sum_{j=1}^{L-1}(\hat{\sigma}_{j}^{x}\hat{\sigma}_{j+1}^{x}+\hat{\sigma}_{j}^{y}\hat{\sigma}_{j+1}^{y})
\end{align}
where $\hat{\sigma}_{j}^{k}$ ($k=x,\ y,\ z$) are the Pauli operators with $j$ labeling the spins in the chain;
$\lambda$ denotes the longitudinal magnetic field exerted homogeneously on all the spins;
$J$ is the coupling strengths between the nearest-neighbor spins of the chain. Set the spin at the first site as subsystem A, 
a time-dependent perturbation of the following form is applied on this subsystem, 
\begin{align}
  F(t)\hat{H}_{1}(t)=F(t)\frac{\hat{\sigma}_{1}^{z}}{2}
\end{align} 
then the total Hamiltonian is given by $\hat{H}(t)=\hat{H}_{0}+\hat{H}_{\text{source}}(t)$.
Diagonalizing $\hat{H}_{0}$ in the single-excitation subspace, its eigenstates read
\begin{align}
    |\varphi_{k}\rangle=\sum_{j=1}^{L}\frac{e^{i(2\pi/L)kj}}{\sqrt{L}}\hat{\sigma}_{j}^{+}|\{\downarrow_{j}\}\rangle
\end{align}
which is a spin wave with wave vector $k$, and its corresponding eigenenergies are $E_{k}=\lambda+2J\cos k$. 
Here $|\{\downarrow_{j}\}\}$ is the ferromagnetic state of the chain with all its spin pointing to the $-\hat{e}_{z}$ direction 
and $\hat{\sigma}_{j}^{+}=(\hat{\sigma}_{j}^{x}+i\hat{\sigma}_{j}^{y})/2$. Any state of the system can be expressed as 
the superposition of these eigenstates 
\begin{align}
    |\psi(t)\rangle=\sum_{k=1}^{L}c_{k}(t)|\varphi_{k}\rangle
\end{align}
where $c_{k}(t)=\langle\varphi_{k}|\psi(t)\rangle$. In  numerical calculation, we choose $c_{k}(t)=\delta_{1,k}$ i.e. $|\psi(t_{0})\rangle=|\varphi_{1}\rangle$, and $F(t)=\alpha\cos(2\pi t/T)e^{-t^{2}/2}$ with $\alpha$ a small prefactor and $JT=\pi$.

\begin{figure}
    \centering
    \includegraphics[width=0.45\textwidth]{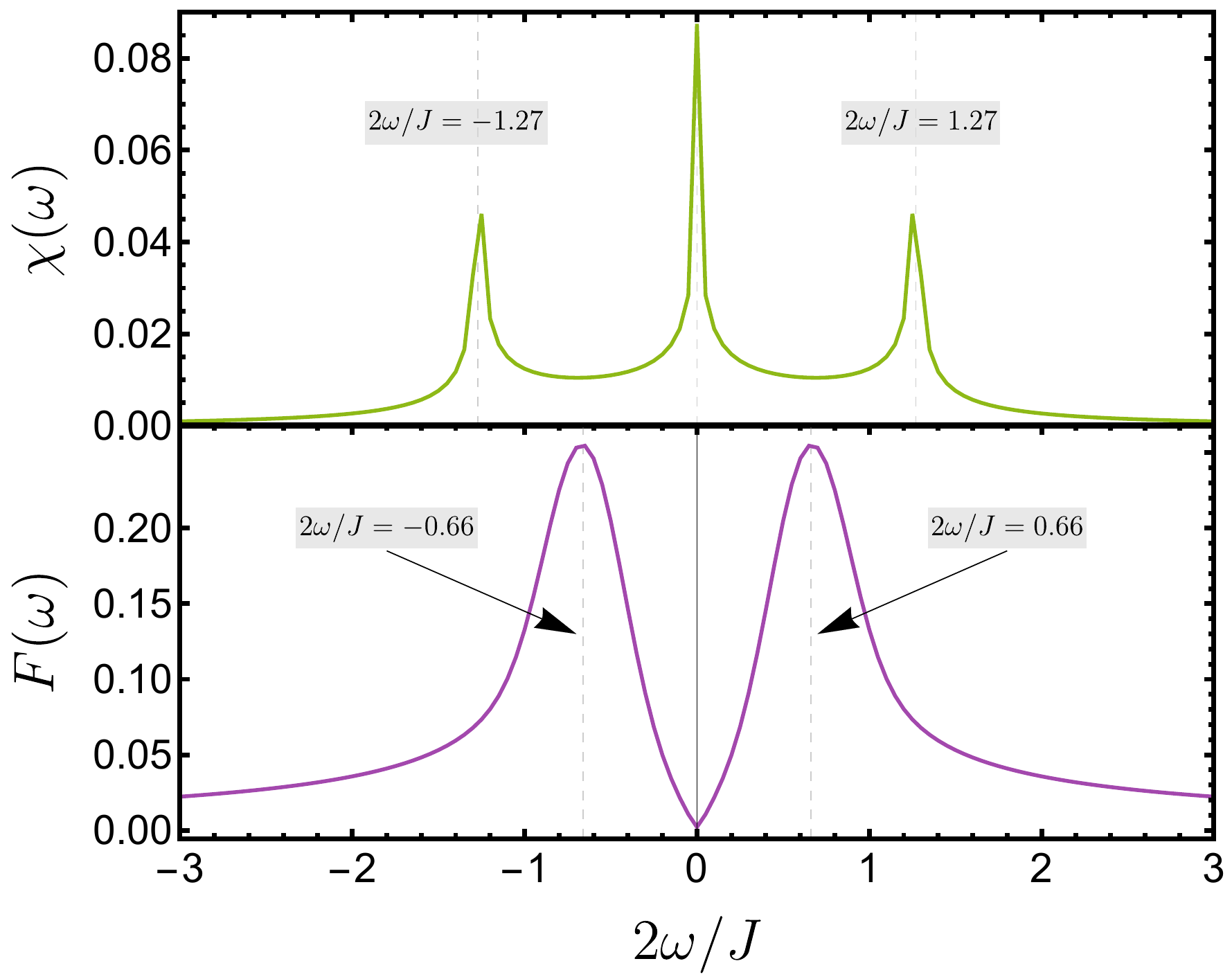}
    \caption{The spectrum $\chi(\omega)$ and $F(\omega)$, which is given by the discrete Fourier transform of response function $R_{E}(t)$ and driving $F(t)$.}
    \label{fig:spectrum}
\end{figure}

Through numerically solving the dynamics of XX spin chain in single-excitation subspace, we obtain $\delta S_{A}(t)$, its spectrum $\delta S_{A}(\omega)$ is given by the discrete Fourier transform of numerical data. 
In Fig.\ref{comp1} and Fig.\ref{comp2}, we show $\delta S_{A}(\omega)$ of the numerical result (blue line) and analytical one which only consider linear correction of dynamics (yellow line). 
There are two characteristic peaks in analytical result, to explain where do these peaks come from, we show the spectrum of $R_{E}(t)$ and $F(t)$ in Fig.\ref{fig:spectrum}, it is readily to see that the one near position $2\omega/J \approx \pm 1.27$ result from the spectrum of response function $\chi(\omega)$; the other peaks near $2\omega/J \approx \pm 0.66$ result from the spectrum of driving $F(\omega)$. Comparing results with different $\alpha$, it is found that when the driving is weak, i.e. $\alpha$ is small enough, numerical results match the linear response quite well, while $\alpha$ became large enough, there is a fairly obvious deviation between numerical spectrum and analytical one, especially at a position near $2\omega/J=0$, $2\omega/J= 2$ and position near $2\omega/J= 2.5$, these deviations result from the emergent of non-linear response.  

\section{\label{sec:conclusion}Conclusion}
An LRT of von Neumann entropy is developed for an open quantum system A which is subject to weak perturbation, 
we assumed A is a subsystem of composite system $A\cup B$, and the composite system is initially at a state which density matrix is commuted with the unperturbed Hamiltonian. 
We found that the LRT of von Neumann entropy to external perturbation is given by the linear response of an observable, using this observable we define the Kubo formula of EE which has the same properties of its conventional counterpart; 
using the LRT of EE, we demonstrate that there is no linear response of von Neumann entropy for a non-degenerate system which is initially at separable states and maximally entangled states. 
To verify our analytical results, 
we numerically solve the dynamics of EE for a subsystem of a $XX$ spin chain. Numerical results show that when the perturbation is weak enough, the dynamics of EE will approach the result given by LRT. 
The LRT of EE provides a useful tool in understanding and investigating the dynamics of entanglement.


\section{\label{sec:ack}acknowledgments}

We acknowledge helpful discussions with Han-Tao Lu and Chong Chen. This work was supported by 
the Natural Science Foundation of Guangdong Province (Grant No.2017B030308003), 
the Key R\&D Program of Guangdong province (Grant No. 2018B030326001), 
the Science, Technology and Innovation Commission of Shenzhen Municipality (Grant No.JCYJ20170412152620376 and No.JCYJ20170817105046702 and No.KYTDPT20181011104202253), 
the National Natural Science Foundation of China (Grant No.11875160, No.U1801661 and No.11905100),
the Economy, Trade and Information Commission of Shenzhen Municipality (Grant No.201901161512), 
the Guangdong Provincial Key Laboratory(Grant No.2019B121203002).


\appendix
\section{\label{derivative_1}Derivation of Eq. (\ref{dyna_state})}

In interaction picture, we formally integral the Liouville equation and obtain
\begin{align}
\rho_{I}(t)=\rho_{0}-i\int_{t_{0}}^{t}d\tau[F(\tau)\hat{H}_{1,I}(\tau),\rho_{I}(\tau)] \label{temp-1-1}
\end{align}
where $\rho_{0}\equiv\rho(t_{0})$, $\hat{H}_{1,I}(t)=\hat{U}_{0}^{\dagger}(t,t_{0})\hat{H}_{1}\hat{U}_{0}(t,t_{0})$ 
and $\rho_{I}(t)=\hat{U}_{0}^{\dagger}(t,t_{0})\rho(t)\hat{U}_{0}(t,t_{0})$, 
this expression is accurate. 
Assuming the value of $F(t)$ is very small, we can expand $\rho_{I}(t)$ in powers of $F(t)$: $\rho_{I}(t)=\rho_{I}^{(0)}(\tau)+F\rho_{I}^{(1)}(\tau)+\mathcal{O}(F^{2})$, 
inserting the expansion into \eqref{temp-1-1}, we obtain 
\begin{align}
\rho_{I}(t)=\rho_{0}-i\int_{t_{0}}^{t}d\tau[F(\tau)\hat{H}_{1,I}(\tau),\rho_{0}]+\mathcal{O}(F^{2}), 
\label{dyna_int}
\end{align}
in the derivation of the above equation, we have used $\rho_{I}^{(0)}(t)=\rho(t_{0})\equiv \rho_{0}$.
we further assume the initial state $\rho_{0}$ commutes with $\hat{H}_{0}$, that is $[\rho_{0},\hat{H}_{0}]=0$.
After transforming the density matrix in the above equation into Schr\"odinger picture, we have
\begin{align}
 \delta \rho(t)=-i\int_{t_{0}}^{t}d\tau[\hat{H}_{1,I}(\tau-t),\rho_{0}]F(\tau)+\mathcal{O}(F^2) \nonumber
\end{align}
where $\delta\rho(t)=\rho(t)-\rho(t_{0})$, we thus get Eq. \eqref{dyna_state}.

\section{\label{derivative_2}Derivation of Eq. (\ref{convolution})}
Working in the interaction picture, we insert Eq. \eqref{dyna_state} into Eq. \eqref{dyna_s}, to the first order in $F(t)$, we obtain the following equation
\begin{align}
    &\delta S_{A}(t) \nonumber\\
    \approx&-i\int_{t_{0}}^{t}d\tau F(\tau)\text{tr}_{A}\left\{\hat{s}_{A}\text{tr}_{B}\left(\big[\hat{H}_{1,I}(\tau-t),\rho_{0}\big]\right)\right\} \nonumber\\
    \approx& -i\int_{t_{0}}^{t}d\tau F(\tau)\text{tr}\left\{\hat{s}_{A}\big[\hat{H}_{1,I}(\tau-t),\rho_{0}\big]\right\} 
\end{align}
where
$\hat{s}_{A}\equiv-\ln \rho_{A}(t_{0})$. Using the cyclic property of trace $\text{tr}(XY)=\text{tr}(YX)$, 
we rewrite the above equation as
\begin{align}
    \delta S_{A}(t)= -i\int_{t_{0}}^{t}\langle \big[\hat{s}_{A},\hat{H}_{1,I}(\tau-t)\big]\rangle_{0} F(\tau)d\tau +\mathcal{O}(F^2)
\end{align}
denoting $R_{E}(t,\tau)=-i\theta(t-\tau)\langle[\hat{s}_{A},\hat{H}_{1,I}(\tau-t)]\rangle_{0}$, the above equation can be simplified into
\begin{align}
    \delta S_{A}(t)\approx\int_{t_{0}}^{\infty}R_{E}(t,\tau)F(\tau)d\tau+\mathcal{O}(F^2)
\end{align}
we thus get Eq. (\ref{convolution}).

It is worth noting that the Kubo formula $R_{E}(t,\tau)$ is invariant under time translations,  that is, $R_{E}(t,\tau)=R_{E}(t-\tau)$.
\section{\label{proof_2}No linear response for maximally entangled states}

Starting from Eq. \eqref{dyna_state}, and assume $[\rho_{0},\hat{H}_{0}]=0$, we may obtain the change in the time evolution state of subsystem A:
\begin{align}
  \delta \rho_{A}(t)&=\text{tr}_{B}[\delta\rho(t)]\nonumber \\
  &=-i\int_{t_{0}}^{t}d\tau\text{tr}_{B}\left([\hat{H}_{1,I}(\tau-t),\rho_{0}]\right)F(\tau) +\mathcal{O}(F^2)
\end{align}
inserting the above equation into $\delta S_{A}(t)= -\text{tr}_{A}[\delta\rho_{A}(t)\ln\rho_{A}(t)]$, we have
\begin{align}
&\delta S_{A}(t) \nonumber\\
& = -i\int_{t_{0}}^{t}d\tau\text{tr}_{A}\left\{ \text{tr}_{B}\left([\hat{H}_{1,I}(\tau-t),\rho_{0}]\right)\hat{s}_{A}\right\}+\mathcal{O}(F^2).
\label{entropy_approx}
\end{align}
Consider a quantum system which consists of two subsystems A and B, the
degree of freedom of A and B are both $d$, assuming initially the total system is at
maximally entangled state: $\rho_{0}=\frac{1}{d}\sum_{i,j=1}^{d}|i_{A}i_{B}\rangle\langle j_{A}j_{B}|$, where $\{|i_{A}\rangle\}$ and $\{|j_{B}\rangle\}$ are complete sets of orthonormal basis of subsystem A and B respectively, we then have
\[
\rho_{A}(t_{0})=\text{tr}_{B}\left(\rho_{0}\right)=\frac{1}{d}\sum_{i}|i_{A}\rangle\langle i_{A}|=\frac{\mathbb{I}_{A}}{d}
\]
then $\ln\rho_{A}(t_{0})=-\mathbb{I}_{A}\ln d$, which is diagonal. On the other hand,  
\begin{align}
 &  \text{tr}_{B}\left([\hat{H}_{1,I}(\tau-t),\rho_{0}]\right)\nonumber\\
 =& \sum_{i,j=1}^{d}\big[\hat{H}_{1,I}^{(j_{B}i_{B})}(\tau-t)|i_{A}\rangle\langle j_{A}|-\hat{H}_{1,I}^{(i_{B}j_{B})}(\tau-t)|j_{A}\rangle\langle i_{A}|\big]
 \label{trace}
\end{align}
where $\hat{H}_{1,I}^{(j_{B}i_{B})}(\tau-t)\equiv \langle j_{B}|\hat{H}_{1,I}(\tau-t)|i_{B}\rangle$. 
It is not hard to see that the above operator is hollow with respect to basis $\{|i_{A}\rangle\}$, therefore according to Eq.\eqref{trace}, we conclude that the response of EE to first order in $\alpha$ is zero: $\delta S_{A}(t)\approx 0$.

\bibliography{ref_LR}

\end{document}